\documentclass[12pt]{article}
\usepackage{amssymb,amsmath,cite}
\usepackage[dvips]{graphicx,color}
\usepackage{psfrag,subfigure}
\usepackage{booktabs}
\usepackage{hyperref}
\title{Calculation of the Vacuum Energy Density using \; \;\; \; Zeta Function Regularization}
\author{Siamak Tafazoli\footnote{\, \copyright \, Siamak Tafazoli \; \; \; email: siamak.tafazoli@ronininstitute.org}  }

\date{}

\begin{document}

\maketitle

\begin{quotation}
\noindent
\small
{\bf Abstract} - This paper presents a theoretical calculation of the vacuum energy density by summing the contributions of all quantum fields' vacuum states which turns out to indicate that there seems to be a missing bosonic contribution in order to match the predictions of current cosmological models and all observational data to date. The basis for this calculation is a Zeta function regularization method used to tame the infinities present in the improper integrals of power functions. The paper also makes a few other contributions in the area of vacuum energy.

\end{quotation} 

\normalsize
 
Much has been written on the \textit {Cosmological Constant (CC) problem} and the \textit {vacuum catastrophe} where we have the worst discrepancy between theory and measurement of about a factor of more than 120 orders of magnitude, refer to \cite{Carroll-2000} for a pedagogical overview.  In this paper, we improve this discrepancy by more than half which still remains very high, but also propose 3 solutions, one of which is an exciting prediction that can help resolve the vacuum catastrophe. We attack the heart of the problem which is the divergent sums and integrals that have also plagued many other areas of Quantum Field Theory (QFT), using as our tool, a Zeta regularization technique to tame these infinities \cite{Aghili-Tafazoli-2018}. Refer also to \cite{Elizalde-2021} for a pedagogical overview of the Zeta function and the associated Zeta function regularization techniques with applications to physics.
\\

The total vacuum (zero-point) energy of a free (non-interacting) quantum field in flat spacetime, can be modeled as the sum of zero-point energies of a set of infinite number of quantum harmonic oscillators, one for each normal mode and given simply as ${+ \sum_{k} \frac{1}{2} \hbar \omega_{k}}$ for bosonic fields and ${- \sum_{k} \frac{1}{2} \hbar \omega_{k}}$ for fermionic fields \cite{Pauli-1971}, \cite{Klauber-2013}, and where ${\frac{1}{2} \hbar \omega_{k}}$ represents the eigenvalues of the free Hamiltonian, ${\omega_{k}= \sqrt{m^2+k^2}}$ in Natural units, ${k}$ is the wave number (with units of \text {1/Length}), ${m}$ is the particle mass associated with a specific field, and ${\hbar}$ is the reduced Planck constant. The negative sign, in front of the infinite sum, is due to the negative energy solutions allowed in the fermionic fields which must follow the Pauli Exclusion principle by obeying the canonical anti-commutation relations.
 
Consequently, the vacuum energy density of a single state of a given field (i.e. energy ${E}$ per volume ${V}$) can be derived \cite{Pauli-1971}, \cite{Visser-2018}, \cite{Klauber-2013} (section 10.8) to be 

\begin{equation} \label{eq:ZPE_def}
\hat{\rho}_{vac}=   \frac{E}{V} = \pm \frac{1}{V}  \sum_{k} \frac{1}{2} \hbar \omega_{k} = ... =
\pm \frac{1}{4\pi^2} \int_{0}^\infty  \sqrt{m^2+k^2} k^2 {dk}
\end{equation} 

and by reintroducing ${\hbar}$ and ${c}$ in \eqref{eq:ZPE_def}, the total vacuum energy density of a field is given by 

\begin{equation} \label{eq:ZPE_def0}
\hat{\rho}_{vac}=
g\frac{1}{4\pi^2} \int_{0}^\infty  \sqrt{(m c^2)^2+(\hbar kc)^2} k^2 {dk}
\end{equation} 

where $g$ is the degeneracy factor which includes, a sign factor $(-1)^{2j}$, a spin factor $2j+1$ for massive fields and 2 for massless fields, a factor of 3 for fields with color charge, and a factor of 2 for fields that have an antiparticle distinct from their particle. Note also that ${c}$ is the speed of light in vacuum and ${j}$ is the spin. It is also important to note that Pauli \cite{Pauli-1971} introduced the following 3 polynomial-in-mass and 1 logarithmic-in-mass conditions in order to obtain a vanishing vacuum energy density:

\begin{equation} \label{eq:ZPE_con}
\sum_{i} g_{i}m_{i}^4 = 0 \; \; \; \; \; \; \sum_{i} g_{i}m_{i}^2 = 0 \; \; \; \; \; \; \sum_{i} g_{i} = 0 \; \; \; \; \; \; \sum_{i} g_{i}m_{i}^4 \ln \Big( \frac{m_{i}^2}{\mu^2}\Big)= 0
\end{equation}
\\
where index $i$ represents different particle species and $\mu$ is an arbitrary parameter.  We will show that using our zeta regularization method, we can simplify this by eliminating 3 of the conditions, leaving only the key quartic polynomial-in-mass condition.
\\

Equation \eqref{eq:ZPE_def0} can be written as 

\begin{align} 
\notag \hat{\rho}_{vac}=  g\frac{\hbar c}{4\pi^2}
\int_{0}^\infty  \sqrt{\Big(\frac{mc}{\hbar}\Big)^2+k^2} k^2 {dk} 
=g\frac{\hbar c}{4\pi^2} \int_{0}^\infty  \sqrt{a^2+k^2} k^2 {dk} \\
\label{eq:ZPE_def1}
=g\frac{\hbar c}{4\pi^2}
\int_{0}^\infty a \sqrt{1+\Big(\frac{k}{a}\Big)^2} k^2 {dk}
\end{align} 
\\
where ${a}=\frac{mc}{\hbar}$  (with units of \text {1/Length}). Now, by introducing a dimensionless change of variable, ${x = k/a}$ and ${dx = dk/a}$, we can turn the integral in \eqref{eq:ZPE_def1} into a `pure' integral and hence ready for Zeta function regularization:

\begin{align} 
\notag \hat{\rho}_{vac}=  g\frac{\hbar c}{4\pi^2} \int_{0}^\infty  a \sqrt{1+x^2} (ax)^2 a{dx} =  g\frac{\hbar c}{4\pi^2} a^4 \int_{0}^\infty  \sqrt{1+x^2} x^2 {dx}  \\
\label{eq:ZPE_zfr}
=g\frac{m^4c^5}{4\pi^2 \hbar^3} \int_{0}^\infty  \sqrt{1+x^2} x^2 {dx} 
\end{align}

In order to apply the Zeta regularization results obtained in appendix A, we need to perform a Maclaurin series expansion of the integrand in equation \eqref{eq:ZPE_zfr} with respect to the dimensionless parameter ${x}$. Let's define 
 
\begin{equation} \label{eq:ZPE_fk}
f(x)= x^2 \sqrt{1+x^2} 
\end{equation}

Now the Maclaurin series expansion of \eqref{eq:ZPE_fk} can be obtained as 

\begin{equation} \label{eq:ZPE_fk2}
f(x)=  x^2 + \frac{x^4}{2} - \frac{x^6}{8}  + \frac{x^8}{16}  - \frac{5x^{10}}{128} + \frac{7x^{12}}{256} - \frac{21x^{14}}{1024} + O(x^{16}, x^{18}, x^{20},...) 
\end{equation}
\\
At first glance, when substituting \eqref{eq:ZPE_fk2} as the integrand in \eqref{eq:ZPE_zfr}, the integral in \eqref{eq:ZPE_zfr} looks highly divergent, essentially a sum of divergent even powered polynomial integrals. However, finite values for these integrals can be assigned using a Zeta regularization technqiue (see appendix A).  These integral values are ${\int_{0}^\infty x^2 dx}={-\frac{1}{12}}$,  ${\int_{0}^\infty x^4 dx}={-\frac{1}{30}}$,  ${\int_{0}^\infty x^6 dx}={-\frac{1}{56}}$ and  ${\int_{0}^\infty x^8 dx}={-\frac{1}{90}}$, etc. An approximate finite zeta-regularized value can hence be obtained for the divergent integral in \eqref{eq:ZPE_zfr} as $\int_{0}^\infty  \sqrt{1+x^2} x^2 {dx}  \approx - \frac{1}{10}$ and so we can finally get the following simple expression for \eqref{eq:ZPE_zfr}

\begin{equation} \label{eq:ZPE_def}
\hat{\rho}_{vac} = - g\frac{m^4c^5}{40\pi^2 \hbar^3} \equiv - g\frac{m^4}{40\pi^2} \;\; (Natural \; Units)
\end{equation}
\\
Now, the sign of the vacuum energy density in \eqref{eq:ZPE_def0} is represented by the $(-1)^{2j}$ sign factor which dictates that the vacuum energy contribution is positive for bosonic fields (e.g. $j=1$) and negative for fermionic fields (i.e. $j=1/2$), but only if one assumes that the integral in \eqref{eq:ZPE_def0} is a positive number tending to plus infinity (i.e. we obtain a positive value for the integral if we use an Ultraviolet (UV) cut-off which is what is typically done in literature). However, this is not what we obtain using our Zeta regularization technique where we assign a \textit{negative} value to the otherwise divergent integral $\int_{0}^\infty  \sqrt{1+x^2} x^2 {dx}$ in \eqref{eq:ZPE_zfr}, and by doing so we reverse the vacuum energy density signs for the fermionic fields (positive) and bosonic fields (negative). This is indeed an unexpected turn of events!
\\

The vacuum energy density contributions of the relativistic massless photon and gluon particles are considered to be zero and hence will not be further considered in the calculations that follows. This can also be concluded from the equation of state for a relativistic particle \cite{Gueorguiev-2020} (page 8).
\\

The total expectation value of the vacuum energy density can be obtained by summing over the free massive quantum fields of the Standard Model (represented by the index $i$), with their associated particle masses \cite{Zyla-2020}:

\begin{equation} \label{eq:ZPE_tot}
\hat{\rho}_{total}=  \sum_{i} (\hat{\rho}_{vac})_{i} = -\frac{1}{40\pi^2}  \sum_{i} g_{i}m_{i}^4
\end{equation}

from which we can clearly conclude now that the only condition required to have a vanishing $\hat{\rho}_{total}$ is the first one in \eqref{eq:ZPE_con}, $\sum_{i} g_{i}m_{i}^4 = 0$.
\\ \\
The vacuum energy density values are tabulated below where the leptons (x6) and quarks (x6) masses are lumped together.  We note that the total expectation value of the vacuum energy density \textit{does not diverge}, is positive, but still quite large.  In fact, we obtain a positive value of $ \hat{\rho}_{total} = 5.93\times 10^{44} $ J/m$^3$ (= $2.85\times 10^{7}$ GeV$^4$) and a corresponding CC value of $\Lambda = 4.84 \times 10^{-12}$ eV$^2$ using \cite{Carroll-2000} (Eqn 14)

\begin{equation} \label{eq:lambda}
\rho_{vac} = \frac{\Lambda}{8 \pi G}
\end{equation}
  
where $G$ is Newton's gravitational constant. \\
\\
This result is a significant improvement over the current theoretical estimates which are about 120 orders of magnitude higher compared with the value obtained from the large scale cosmological observations of approximately $ 5.26\times10^{-10} $ J/m$^3$ ($\Lambda = (4.24\pm0.11) \times 10^{-66}$ eV$^2$) \cite{Planck Collaboration2018}. However, considering carefully the order of magnitude of our newly calculated vacuum energy density ($10^{7}$ GeV$^4$), we make a startling observation that we have indeed obtained a value in the right energy scale which is suggestive of at least one missing heavy boson (e.g. a heavy cousin of the Higgs boson). For example, if we assume a scalar boson (with $g= 1$), then using \eqref{eq:ZPE_def}, its mass would be about $327$ GeV/$c^2$. If we assume a vector boson (e.g. say $g=3$ for a Z-prime boson), then its mass would be about $247$ GeV/$c^2$.
\\

We conclude that the results presented in this paper help partly improve the vacuum catastrophe or CC problem by reducing the existing discrepancy of more than 120 order of magnitude to about 55. We also propose 3 possible solutions to explain this discrepancy that still exists between the new theoretical and the observational values which may be wholly or partly due to:
\\
\\
(i) the vacuum energy contribution of at least one unknown massive bosonic particle, as described above. Of course, it is entirely possible that we are also missing a massive fermionic (dark matter) particle which, if detected and of consequential mass, would further push up the predicted mass of the missing boson; \\
(ii) the vacuum energy contribution of the gravitational field or the effect of spacetime curvature on vacuum energy; and \\
(iii) the effect of interacting quantum fields on vacuum energy density.\\ \\

In summary, we list the key contributions of the paper:
\\
\\
1)	Vacuum energy density is finite without the need of a UV cut-off as is typically done in the literature; \\ 
2)	Vacuum energy density is found to have a simple closed-form equation \eqref{eq:ZPE_def} which is a quartic function of the particle rest mass; \\
3)	Only one of Pauli’s original conditions is necessary and sufficient to zero out the vacuum energy density; and \\
4)	The vacuum energy contributions from the fermionic fields are positive and from the bosonic fields are negative.

\begin{table}
\centering
\begin{tabular}{|l|l|l|l|} 
\toprule
Fields                                                    &  \begin{tabular}[c]{@{}l@{}}Degeneracy factor $g$ \\sign $\times$ spin $\times$ color $\times$ antiparticle  \end{tabular} & Mass (eV/$c^2$) & $  \hat{\rho}_{vac}$ (J/m$^3$)  \\ 
\hline
\hline
Quarks~ (x 6): fermionic                                            &  $-1 \times$ (2$\times$1/2+1) $\times$ 3 $\times$ $2 = -12 $   & $178.31 \times 10^{9} $       & $+6.29\times 10^{44} $     \\ 
\hline
Leptons~(x 6): fermionic                                            &  $-1 \times$ (2$\times$1/2+1) $\times$ 1 $\times$ $2 = -4$    & $1.90\times 10^{9} $        & $+2.70\times 10^{36} $     \\ 
\hline
W: bosonic                                                   & 1 $\times$ (2$\times$1+1) $\times$ 1 $\times$ 2 = 6      & $80.38 \times 10^{9} $       & $-1.30\times 10^{43} $     \\ 
\hline
Z : bosonic                                                    &  1 $\times$ (2$\times$1+1) $\times$ 1 $\times$ 1 = 3      & $91.19 \times 10^{9} $       & $-1.08\times 10^{43} $     \\
\hline
Higgs : bosonic                                                      & 1 $\times$ (2$\times$0+1) $\times$ 1 $\times$ 1 = 1     & $ 125.10 \times 10^{9} $      & $-1.27\times  10^{43} $     \\ 
\hline
Total                                                 &        &                &      $+5.93\times 10^{44} $        \\

\bottomrule
\end{tabular}
\end{table}

\bibliographystyle{IEEEtran} 

\appendix
\section{Calculation of the Vacuum Energy Density using Zeta Function Regularization}
Let's define the $\mu$ function as the improper integral of a power function with exponent $p$  
\begin{equation} \label{eq:Ip_def}
\mu(p)=\int_0^{\infty} x^p dx 
\end{equation}
with $p \in \mathbb{N}^{0}$, where  $\mathbb{N}^{0}=\{0, 1, 2, \cdots \} $.
\\

The mu function can be interpreted as the natural extension of the Riemann zeta function where the discrete sum is replaced by a continuous integral. The above improper integral can be equivalently written as the following infinite series by splitting the limits of integration into successive integer numbers. That is   
\begin{subequations}
\begin{equation} \label{eq:sp=sumDelta}
\mu(p)=\sum_{n=1}^{\infty} \Delta(p,n)
\end{equation} 
where 
\begin{equation} \label{eq:Delta_p}
\Delta(p,n)=\int_{n-1}^n x^p dx =\frac{1}{p+1}\Big( n^{p+1} - (n-1)^{p+1}  \Big)
\end{equation}
\end{subequations}
is the definite integral of the power function over limits $n-1$ and $n$. In what follows, we will show that the divergent series in the RHS of \eqref{eq:sp=sumDelta} can converge to a finite result!
\\

One can write the binomial  polynomial expansion of $(n-1)^{p+1}$ in \eqref{eq:Delta_p} as \cite{Wolfram-BT} (Eqn 2)
\begin{subequations}
\begin{equation} \label{eq:n-1_binomial}
(n-1)^{p+1} = \sum_{k=0}^{p+1} \binom{p+1}{k} n^k (-1)^{p+1-k},
\end{equation}
where
\begin{equation} \label{binom}
\binom{t}{s} = \frac{t!}{s !(t-s)!}	
\end{equation}
\end{subequations}
is  the binomial  coefficient, with $\binom{t}{t}=\binom{t}{0}=1$.  Substituting  \eqref{eq:n-1_binomial} into \eqref{eq:Delta_p} gives
\begin{align} 
\notag \Delta(p,n) &=\frac{1}{p+1} \Big( n^{p+1} - \sum_{k=0}^{p+1} \binom{p+1}{k} n^k (-1)^{p+1-k} \Big) \\
\notag &= \frac{1}{p+1} \Big( n^{p+1} - \sum_{k=0}^{p} \binom{p+1}{k} n^k (-1)^{p-k} (-1)^{1} - \binom{p+1}{p+1} n^{p+1} (-1)^{p+1-(p+1)}\Big) \\
\label{eq:Delta_p1}
&= \frac{1}{p+1}\sum_{k=0}^{p} \binom{p+1}{k} n^k(-1)^{p-k}
\end{align}
Thus \eqref{eq:sp=sumDelta} can be written as
\begin{align} \notag 
\mu(p) & =\frac{1}{p+1}\sum_{n=1}^{\infty} \sum_{k=0}^{p} \binom{p+1}{k} (-1)^{p-k}n^k \\ \label{eq:Delta_p2}
& = \frac{1}{p+1}  \Big( \sum_{k=0}^{p} \binom{p+1}{k}(-1)^{p-k}   
\sum_{n=1}^{\infty} n^k \Big) \qquad \forall p \in \mathbb{N}^{0} 
\end{align}
The last summation in RHS of \eqref{eq:Delta_p2} can be formally written  
in terms of  the Reimann zeta function $\zeta(\cdot)$ as
\begin{equation} \label{eq:zeta}
\zeta(-k) = \sum_{n=1}^{\infty} n^{k}
\end{equation}
and therefore we arrive at
\begin{equation} \label{eq:sp}
\mu(p)=\frac{1}{p+1}\sum_{k=0}^{p} \binom{p+1}{k} (-1)^{p-k} \zeta(-k)   \qquad \forall p \in \mathbb{N}^{0}
\end{equation}
For integers $k\geq 0$, the zeta function is related to Bernoulli numbers by \cite{Wolfram-Zeta} (Eqn 64)
\begin{equation} \label{eq:zeta_B}
\zeta(-k) = (-1)^k \frac{B_{k+1}}{k+1}
\end{equation}
Moreover, from definition \eqref{binom}, one can verify that the successive binomial coefficients hold the following useful identity   
\begin{equation} \label{eq:2Cs}
\binom{p+1}{k}  = \frac{k+1}{p+2} \binom{p+2}{k+1}
\end{equation}
Finally, upon substituting the relevant terms from \eqref{eq:zeta_B} and \eqref{eq:2Cs} into \eqref{eq:sp}, one can equivalently rewrite the latter equation in the following simple form 

\begin{align} 
\notag \mu(p)&=\frac{1}{p+1}\sum_{k=0}^{p} \frac{k+1}{p+2} \binom{p+2}{k+1} (-1)^{p-k} (-1)^k \frac{B_{k+1}}{k+1}   \\
\notag &=\frac{(-1)^{p}}{(p+1)(p+2)}\sum_{k=0}^{p} \binom{p+2}{k+1} B_{k+1}   \\ \label{eq:Sp2}
&= \frac{(-1)^{p}}{(p+1)(p+2)}\sum_{k=1}^{p+1} \binom{p+2}{k} B_{k} \qquad \forall p \in \mathbb{N}^{0}
\end{align}

Furthermore, the Bernoulli numbers satisfy the following property \cite{Wolfram-Bernoulli} (Eqn 34)
\begin{equation} \label{eq:sumCB1}
\sum_{k=0}^{l-1} \binom{l}{k} B_k =0
\end{equation}
Given that $B_0=1$ and using a change of variable $l=p+2$, one can equivalently write  \eqref{eq:sumCB1} as 
\begin{align} 
\notag \sum_{k=1}^{p+1} \binom{p+2}{k} B_k + \binom{p+2}{0} B_0 = 0 \\
\label{eq:sumCB2}
\sum_{k=1}^{p+1} \binom{p+2}{k} B_k = -1
\end{align}
Finally upon substituting  \eqref{eq:sumCB2} in \eqref{eq:Sp2}, we arrive at the explicit expression of the $\mu$ function  
\begin{equation}  \label{eq:Sp3}
\boxed {\mu(p)=\int_0^{\infty} x^p dx =\frac{(-1)^{p+1}}{(p+1)(p+2)}  \qquad \forall p \in \mathbb{N}^{0} }
\end{equation}
\\
This surprising and simple finite result should be viewed as an alternative, that may be useful in certain cases, to an otherwise divergent solution! \\ \\

As a simple check of \eqref{eq:Sp3} for small $p$ (e.g. $p=0,1,2$), and starting from  \eqref{eq:sp=sumDelta} and \eqref{eq:Delta_p}:

\begin{align} 
\notag \mu(p)=\int_0^{\infty} x^p dx = \sum_{n=1}^{\infty}\int_{n-1}^n x^p dx=\sum_{n=1}^{\infty}\frac{1}{p+1}\Big( n^{p+1} - (n-1)^{p+1}  \Big)
\end{align}
\\
\\
we get for $p=0$:
\begin{align} 
\notag \mu(0)=\int_0^{\infty} x^0 dx = \sum_{n=1}^{\infty}\int_{n-1}^n 1 dx=\sum_{n=1}^{\infty} ( n - (n-1) )=\sum_{n=1}^{\infty} 1=\sum_{n=1}^{\infty} n^{0}=\zeta(0)=-\frac{1}{2}= \frac{(-1)^{0+1}}{(0+1)(0+2)}
\end{align}
\\
\\
for $p=1$:
\begin{align} 
\notag \mu(1)&=\int_0^{\infty} x^1 dx = \sum_{n=1}^{\infty}\int_{n-1}^n x dx=\sum_{n=1}^{\infty}\frac{1}{2}\Big( n^{2} - (n-1)^{2}  \Big)= \frac{1}{2}\sum_{n=1}^{\infty}\Big( n^{2} - (n^{2}-2n+1)  \Big) \\
\notag &= \frac{1}{2}\sum_{n=1}^{\infty}(n^{2} - n^{2}+2n-1)=\frac{1}{2}\sum_{n=1}^{\infty}(2n-1)=\frac{1}{2}(2\zeta(-1)-\zeta(0))\\
\notag &= \frac{1}{2}\Big(2 \Big(\frac{-1}{12}\Big)-\frac{-1}{2}\Big)= \frac{1}{2}\Big(\frac{-1+3}{6}\Big)=\frac{1}{6}=\frac{(-1)^{1+1}}{(1+1)(1+2)}
\end{align}
\\
\\
for $p=2$:
\begin{align} 
\notag \mu(2)&=\int_0^{\infty} x^2 dx = \sum_{n=1}^{\infty}\int_{n-1}^n x^2 dx=\sum_{n=1}^{\infty}\frac{1}{3}\Big( n^{3} - (n-1)^{3}  \Big)= \frac{1}{3}\sum_{n=1}^{\infty}\Big( n^{3} - (n^{3}-3n^{2}+3n-1)  \Big) \\
\notag &= \frac{1}{3}\sum_{n=1}^{\infty}(3n^{2}-3n+1)=\frac{1}{3}\Big(3\sum_{n=1}^{\infty}n^{2}-3\sum_{n=1}^{\infty}n^{1}+\sum_{n=1}^{\infty}n^{0}\Big)=\frac{1}{3} (3\zeta(-2)-3\zeta(-1)+\zeta(0))\\
\notag &= \zeta(-2)-\zeta(-1)+\frac{1}{3}\zeta(0) = 0- \Big(\frac{-1}{12}\Big)+\frac{1}{3}\Big(\frac{-1}{2}\Big)= \frac{1}{12}-\frac{1}{6}=-\frac{1}{12}=\frac{(-1)^{2+1}}{(2+1)(2+2)}
\end{align}

\end{document}